\documentclass[prl,twocolumn,superscriptaddress,nofootinbib]{revtex4}

\usepackage{amsfonts}
\usepackage{amsmath}
\usepackage{amssymb}
\usepackage{amsthm}
\usepackage{bm}
\usepackage{dcolumn}
\usepackage{epsfig}
\usepackage{graphicx}
\usepackage{graphics}
\usepackage[latin1]{inputenc}
\usepackage{latexsym}
\usepackage{rotating}
\usepackage{hyperref}

\begin{document}
\title{A Tale of Two Jets}

\author{Nicol\'as Yunes}
\affiliation{Department of Physics, Princeton University, Princeton, NJ 08544, USA.}
\affiliation{Department of Physics and MIT Kavli Institute, MIT, Cambridge, MA 02139, USA.}

\date{\today}

\begin{abstract}
\end{abstract}

\maketitle


One of the most astounding astrophysical phenomena are highly collimated and energetic relativistic jets from a highly localized sky location. These jets are ubiquitous, observed to high redshift in the radio and x-ray spectrum at the center of galaxies, gamma-ray bursts, quasars, etc. Although their precise generation mechanism remains a bit of a mysetery, these jets are believed to be powered by black holes, one of the most striking consequences of Einstein's theory of General Relativity. These incredibly dense objects do not shine in isolation because their gravitational field is so strong that not even light can escape their pull. Black holes, however, are usually accompanied by accretion disks and can channel strongly collimated emissions through a variety of mechanisms. Recently, Palenzuela, Lehner and Liebling~\cite{Palenzuela:2010nf} studied the merger of two black holes, surrounded by a circumbinary accretion disk that anchors a magnetic field perpendicular to the orbital plane. In such a system, each black hole acts as a ``straw'' that stirs the magnetic field around as the black holes inspiral. The twisting of the magnetic field then generates jets around each black hole, even though these are not spinning. Their simulations show the formation of such a dual jet geometry and how it transitions to a single jet one, as the black holes merge due to gravitational wave emission.

The jet formation scenario can be partially understood through the seminal work of Blandford and Znajek~\cite{1977MNRAS.179..433B}. They considered a single spinning black hole with a magnetized accretion disk, supported by external currents flowing in the disk. This disk generates a magnetic field that is dragged by the spinning black hole, inducing an electric field that accelerates stray electrons, forcing them to pair-produce and create a neutral plasma. This plasma is then believed to be funneled by the magnetic field lines, that now rotate with the black hole, leading to synchrotron radiation. In this way, the plasma is believed to generate a highly-collimated electromagnetic flux that extracts rotational energy from the black hole, an scenario that has been confirmed via  numerical simulations~\cite{2004MNRAS.350.1431K,2004Sci...305..978S,2007MNRAS.377L..49K,2008MNRAS.388..551T,2010LNP...794..265K}

Supermassive black holes, however, do not exist in isolation; they have been observed at the center of most galaxies, which in turn have also been observed to undergo mergers with other galaxies at moderate redshifts. As these galaxies merge, the central black holes in each galaxy will sink to the center of the remnant due to dynamical friction and form a binary system. While doing so, the accretion disks of each black hole combine into a circumbinary disk, which anchors a magnetic field in the inner region. This region is essentially devoid of material because the black holes evacuate it as they inspiral due to loss of binding energy by gravitational wave emission. The black hole binary eventually merges, forming a spinning black hole that is expected to possess a Blandford-Znajek jet. Due to the strong non-linearities and complexity of the mathematics that describes this scenario, the standard analytical and perturbative techniques used in Blandford-Znajek-type calculations fail, obscuring the process through which such jets form.

The simulations of Palenzuela, Lehner and Liebling shed light on this problem. Their simulations are the first, fully-relativistic treatment of an inspiraling and merging binary black hole system, coupled to Maxwell equations to implicitly model the behavior of plasma inside the circumbinary disk. Such evolutions are possible thanks to recent advances in numerical relativistic simulations of the late inspiral and merger of two, otherwise isolated black holes in full General Relativity~\cite{Pretorius:2005gq,Campanelli:2005dd,Baker:2006yw}. In their simulations,  Palenzuela, Lehner and Liebling enhanced such binary, non-spinning black hole evolutions through the inclusion of a magnetic field perpendicular to the orbital plane that is evolved via Maxwell's equations coupled to the Einstein equations. The magnetic field is supported by currents in the circumbinary disk, which decouples from the evolutions as it is far from the binary. The electromagnetic field is evolved in a force-free approximation~\cite{1969ApJ...157..869G,1977MNRAS.179..433B}, where the bulk of the plasma is assumed to experience no Lorentz force due to its negligible fluid inertia.

These simulations revealed that, even in the absence of black hole spins, dual jets are induced around each black hole by their orbital motion, and these jets merge into a single Blandford-Znajek one after the black holes merge. Initially, when they are well-separated, the black holes inspiral and stir the surrounding plasma and magnetic field lines. By electromagnetic induction, this motion produces a poloidal electric field, a toroidal magnetic field and a Poynting flux around each black hole, leading to dual jets. Unlike in the Blandford-Znajek mechanism, the black hole are here non-spinning and the jets are powered by the black hole's kinetic energy, instead of their rotational energy. Eventually, these objects merge and form a highly distorted and spinning black hole, at which point a strong isotropic burst of electromagnetic radiation is emitted, with twice the amount of energy relative to the individual jets. In addition, the dual jet geometry transitions to a single, collimated jet, which, after the remnant settles to a spinning black hole, is well-described by the Blandford-Znajek model. These simulations thus prove that Blandford-Znajek jets are incredibly robust and stable as end states of binary evolutions, even in highly dynamical and asymmetric spacetimes. 

Apart from discovering a generalization of the Blandford-Znajek jet formation mechanism and elucidating the fate of dual jets in binary black hole mergers, the simulations of Palenzuela, Lehner and Liebling also have strong implications to gravitational wave astrophysics. This scenario presents itself as a beautiful electromagnetic counterpart to the detection of gravitational waves from binary black hole mergers. Such a counterpart should be electromagnetically detectable as it presents a very clear energy profile. During inspiral, the dual jets lead to an $m=2$ electromagnetic structure, lasting several days and with energies proportional to the magnetic field times the orbital velocity squared. Such emission remains roughly constant over most of the inspiral, as the orbital velocity varies slowly. When the binary merges, a strong isotropic burst of radiation is emitted and the dual jets transition to a Blandford-Znajek one with $m=0$ electromagnetic structure. If such a counterpart is observed, it would provide information on the location of the event in the sky and its redshift. Such an electromagnetic observation could then be used in gravitational wave observations to measure other parameters more accurately, such as the Hubble constant~\cite{Schutz:1986gp}, the dark energy equation of state~\cite{Holz:2005df}, or to test deviations from General Relativity~\cite{Yunes:2010yf}.

\bibliographystyle{apsrev}
\bibliography{review}

\begin{thebibliography}{14}
\expandafter\ifx\csname natexlab\endcsname\relax\def\natexlab#1{#1}\fi
\expandafter\ifx\csname bibnamefont\endcsname\relax
  \def\bibnamefont#1{#1}\fi
\expandafter\ifx\csname bibfnamefont\endcsname\relax
  \def\bibfnamefont#1{#1}\fi
\expandafter\ifx\csname citenamefont\endcsname\relax
  \def\citenamefont#1{#1}\fi
\expandafter\ifx\csname url\endcsname\relax
  \def\url#1{\texttt{#1}}\fi
\expandafter\ifx\csname urlprefix\endcsname\relax\def\urlprefix{URL }\fi
\providecommand{\bibinfo}[2]{#2}
\providecommand{\eprint}[2][]{\url{#2}}

\bibitem[{\citenamefont{Palenzuela et~al.}(2010)\citenamefont{Palenzuela,
  Lehner, and Liebling}}]{Palenzuela:2010nf}
\bibinfo{author}{\bibfnamefont{C.}~\bibnamefont{Palenzuela}},
  \bibinfo{author}{\bibfnamefont{L.}~\bibnamefont{Lehner}}, \bibnamefont{and}
  \bibinfo{author}{\bibfnamefont{S.~L.} \bibnamefont{Liebling}}
  (\bibinfo{year}{2010}).

\bibitem[{\citenamefont{{Blandford} and {Znajek}}(1977)}]{1977MNRAS.179..433B}
\bibinfo{author}{\bibfnamefont{R.~D.} \bibnamefont{{Blandford}}}
  \bibnamefont{and} \bibinfo{author}{\bibfnamefont{R.~L.}
  \bibnamefont{{Znajek}}}, \bibinfo{journal}{MNRAS}
  \textbf{\bibinfo{volume}{179}}, \bibinfo{pages}{433} (\bibinfo{year}{1977}).

\bibitem[{\citenamefont{{Komissarov}}(2004)}]{2004MNRAS.350.1431K}
\bibinfo{author}{\bibfnamefont{S.~S.} \bibnamefont{{Komissarov}}},
  \bibinfo{journal}{MNRAS} \textbf{\bibinfo{volume}{350}},
  \bibinfo{pages}{1431} (\bibinfo{year}{2004}),
  \eprint{arXiv:astro-ph/0402430}.

\bibitem[{\citenamefont{{Semenov} et~al.}(2004)\citenamefont{{Semenov},
  {Dyadechkin}, and {Punsly}}}]{2004Sci...305..978S}
\bibinfo{author}{\bibfnamefont{V.}~\bibnamefont{{Semenov}}},
  \bibinfo{author}{\bibfnamefont{S.}~\bibnamefont{{Dyadechkin}}},
  \bibnamefont{and} \bibinfo{author}{\bibfnamefont{B.}~\bibnamefont{{Punsly}}},
  \bibinfo{journal}{Science} \textbf{\bibinfo{volume}{305}},
  \bibinfo{pages}{978} (\bibinfo{year}{2004}).

\bibitem[{\citenamefont{{Komissarov} and
  {McKinney}}(2007)}]{2007MNRAS.377L..49K}
\bibinfo{author}{\bibfnamefont{S.~S.} \bibnamefont{{Komissarov}}}
  \bibnamefont{and} \bibinfo{author}{\bibfnamefont{J.~C.}
  \bibnamefont{{McKinney}}}, \bibinfo{journal}{MNRAS}
  \textbf{\bibinfo{volume}{377}}, \bibinfo{pages}{L49} (\bibinfo{year}{2007}).

\bibitem[{\citenamefont{{Tchekhovskoy}
  et~al.}(2008)\citenamefont{{Tchekhovskoy}, {McKinney}, and
  {Narayan}}}]{2008MNRAS.388..551T}
\bibinfo{author}{\bibfnamefont{A.}~\bibnamefont{{Tchekhovskoy}}},
  \bibinfo{author}{\bibfnamefont{J.~C.} \bibnamefont{{McKinney}}},
  \bibnamefont{and}
  \bibinfo{author}{\bibfnamefont{R.}~\bibnamefont{{Narayan}}},
  \bibinfo{journal}{MNRAS} \textbf{\bibinfo{volume}{388}}, \bibinfo{pages}{551}
  (\bibinfo{year}{2008}).

\bibitem[{\citenamefont{{Krolik} and {Hawley}}(2010)}]{2010LNP...794..265K}
\bibinfo{author}{\bibfnamefont{J.~H.} \bibnamefont{{Krolik}}} \bibnamefont{and}
  \bibinfo{author}{\bibfnamefont{J.~F.} \bibnamefont{{Hawley}}},
  \textbf{\bibinfo{volume}{794}}, \bibinfo{pages}{265} (\bibinfo{year}{2010}),
  \eprint{0909.2580}.

\bibitem[{\citenamefont{Pretorius}(2005)}]{Pretorius:2005gq}
\bibinfo{author}{\bibfnamefont{F.}~\bibnamefont{Pretorius}},
  \bibinfo{journal}{Phys. Rev. Lett.} \textbf{\bibinfo{volume}{95}},
  \bibinfo{pages}{121101} (\bibinfo{year}{2005}).

\bibitem[{\citenamefont{Campanelli et~al.}(2006)\citenamefont{Campanelli,
  Lousto, Marronetti, and Zlochower}}]{Campanelli:2005dd}
\bibinfo{author}{\bibfnamefont{M.}~\bibnamefont{Campanelli}},
  \bibinfo{author}{\bibfnamefont{C.~O.} \bibnamefont{Lousto}},
  \bibinfo{author}{\bibfnamefont{P.}~\bibnamefont{Marronetti}},
  \bibnamefont{and}
  \bibinfo{author}{\bibfnamefont{Y.}~\bibnamefont{Zlochower}},
  \bibinfo{journal}{Phys. Rev. Lett.} \textbf{\bibinfo{volume}{96}},
  \bibinfo{pages}{111101} (\bibinfo{year}{2006}).

\bibitem[{\citenamefont{Baker et~al.}(2006)\citenamefont{Baker, Centrella,
  Choi, Koppitz, and van Meter}}]{Baker:2006yw}
\bibinfo{author}{\bibfnamefont{J.~G.} \bibnamefont{Baker}},
  \bibinfo{author}{\bibfnamefont{J.}~\bibnamefont{Centrella}},
  \bibinfo{author}{\bibfnamefont{D.-I.} \bibnamefont{Choi}},
  \bibinfo{author}{\bibfnamefont{M.}~\bibnamefont{Koppitz}}, \bibnamefont{and}
  \bibinfo{author}{\bibfnamefont{J.}~\bibnamefont{van Meter}},
  \bibinfo{journal}{Phys. Rev.} \textbf{\bibinfo{volume}{D73}},
  \bibinfo{pages}{104002} (\bibinfo{year}{2006}).

\bibitem[{\citenamefont{{Goldreich} and {Julian}}(1969)}]{1969ApJ...157..869G}
\bibinfo{author}{\bibfnamefont{P.}~\bibnamefont{{Goldreich}}} \bibnamefont{and}
  \bibinfo{author}{\bibfnamefont{W.~H.} \bibnamefont{{Julian}}},
  \bibinfo{journal}{\apj} \textbf{\bibinfo{volume}{157}}, \bibinfo{pages}{869}
  (\bibinfo{year}{1969}).

\bibitem[{\citenamefont{Schutz}(1986)}]{Schutz:1986gp}
\bibinfo{author}{\bibfnamefont{B.~F.} \bibnamefont{Schutz}},
  \bibinfo{journal}{Nature} \textbf{\bibinfo{volume}{323}},
  \bibinfo{pages}{310} (\bibinfo{year}{1986}).

\bibitem[{\citenamefont{Holz and Hughes}(2005)}]{Holz:2005df}
\bibinfo{author}{\bibfnamefont{D.~E.} \bibnamefont{Holz}} \bibnamefont{and}
  \bibinfo{author}{\bibfnamefont{S.~A.} \bibnamefont{Hughes}},
  \bibinfo{journal}{Astrophys. J.} \textbf{\bibinfo{volume}{629}},
  \bibinfo{pages}{15} (\bibinfo{year}{2005}).

\bibitem[{\citenamefont{Yunes et~al.}(2010)\citenamefont{Yunes, O'Shaughnessy,
  Owen, and Alexander}}]{Yunes:2010yf}
\bibinfo{author}{\bibfnamefont{N.}~\bibnamefont{Yunes}},
  \bibinfo{author}{\bibfnamefont{R.}~\bibnamefont{O'Shaughnessy}},
  \bibinfo{author}{\bibfnamefont{B.~J.} \bibnamefont{Owen}}, \bibnamefont{and}
  \bibinfo{author}{\bibfnamefont{S.}~\bibnamefont{Alexander}}
  (\bibinfo{year}{2010}).

\end{thebibliography}

\end{document}